\documentclass[twocolumn]{aa}

\usepackage{amsmath}
\usepackage{amssymb}
\usepackage{graphicx}
\usepackage{dcolumn}
\usepackage{bm}
\usepackage[switch,mathlines]{lineno}

\usepackage{xcolor}

\usepackage{lipsum}
\usepackage{float}

\usepackage{graphicx}

\usepackage{txfonts}

\usepackage{stfloats}

\begin{document}

    \title{Acceleration of relativistic protons in a solar wind perturbed by a coronal mass ejection}

    \author{A. Houeibib \inst{1}
          \and F. Pantellini \inst{1}
          \and L. Griton  \inst{1}
          }

    \institute{LIRA, Observatoire de Paris, Université PSL, Sorbonne Université, Université Paris Cité, CY Cergy Paris Université, CNRS, 92190 Meudon, France}

   \date{......}

    \abstract{
    We investigated the impact of a coronal mass ejection (CME) on the transport and acceleration of relativistic protons in the solar 
    wind using a coupled 3D magnetohydrodynamics (MHD) simulation and a test-particle approach. The CME is driven by a spheromak injected 
    into a Parker solar wind at a heliocentric distance of 0.139 AU. We integrated the trajectories of 5 GeV protons, injected toward the CME from 3 AU,  in the guiding-center approximation and scattered the particles in velocity space with a mean free path $\lambda_{\|}$. 
    Our results show that the CME can increase the protons' energy by several gigaelectronvolts. The acceleration occurs while particles stream 
    along the portion of a magnetic field line downstream of the quasi-perpendicular portion of the CME-driven shock. 
    In our configuration, the maximum energy gain, which is on the order of a few percent per passage through the acceleration region, occurs 
    when the shock approaches 0.3 AU. 
    Large energy gains require multiple passes through the acceleration region, made possible by the combined action of the mirror 
    force and pitch-angle scattering. The efficiency of the acceleration on timescales on the order of hours scales as $\lambda_{\|}^{-3/2}$. 
    Energy spectra harden with decreasing parallel mean free path $\lambda_{\|}$. 
    }


   \maketitle
   \nolinenumbers

\section{Introduction}
At Earth's orbit, the interplanetary medium is mostly filled with protons and electrons at typical thermal energies of $\sim 10$ eV, 
streaming outward from the Sun at several $10^2$ km/s. In addition to these dominant components (in terms of pressure and mass density), the solar wind also contains
a tenuous background of much more energetic particles. These include solar energetic particles (SEPs) accelerated during 
solar eruptions or at  shock waves driven by coronal mass ejections (CMEs) \citep{Desai_2016,Klein_2017}, which reach energies of several 
hundred megaelectronvolts \citep{Reames_1997}. Extra-heliospheric particles, such as anomalous cosmic rays (ACRs) and galactic cosmic rays (GCRs), cover 
an even larger range from a few megaelectronvolts up to several petaelectronvolts  (see e.g., \cite{Lara_etal_2024} and references therein). 
Charged particles are sensitive to the electromagnetic field. Hence, modifications of the interplanetary electromagnetic structure by transients 
such as CMEs are expected to affect the transport of energetic particles throughout the heliosphere (see e.g., \cite{Cane_2000,Richardson_2004,Richardson_2011}).
Many studies on the propagation of GCRs in the interplanetary medium focused on the so-called ``Forbush decrease''  
\citep{Forbush_1937,Forbush_1938,Forbush_1958}, which is a sudden drop in GCR intensity in the wake of a CME \citep{Cane_2000,Dumbovic_2012,Kilpua_2017}. 
Fewer studies have addressed the energy gain or loss of GCRs as they interact with CMEs. 
One reason is that onboard instruments measure GCR flux variations in selected energy channels rather than changes in the kinetic energy of individual GCRs.
In addition, spacecraft can only monitor temporal variations at a single point in space. A more global view can be obtained by studying the propagation 
of GCRs in the electromagnetic field of a 3D, time-dependent numerical simulation of a CME. In this work, we integrated the trajectories of individual  
protons in the guiding-center approximation (GCA),  as in a previous study of 81 keV electrons in a steady solar wind \citep[see][]{houeibib_2025}. 
We extracted the background electromagnetic field from a 3D magnetohydrodynamic (MHD) simulation of a Parker-type solar wind perturbed by a single CME and we considered the propagation of relativistic 5 GeV protons, for which the GCA remains valid in the CME environment near 1 AU. We discuss in detail the role of 
pitch-angle scattering by small-scale turbulence (not included in the MHD simulation).
A popular complementary approach based on the focused transport equation (FTE) has been widely used to investigate the evolution of the velocity distribution 
function of energetic particles \citep[e.g.,][]{Zhang2006,leRoux2009,leRoux2012,vandenBerg2020}. The deterministic momentum-advection term of the FTE 
contains the energy changes associated with the gradient and curvature drifts in the motional electric field. Thus, when the phase-space diffusion coefficients 
are set to zero, the FTE description is equivalent to the guiding-center equations used here. The two approaches differ in practice: FTE-based studies generally focus 
on the structure of the diffusion coefficients for given models of the underlying turbulence, whereas the trajectory-based GCA approach adopted here allows us to 
follow individual particles through the full 3D, time-dependent electromagnetic field of the MHD simulation. As shown by, for example, \citet{Marsh_etal_2013} and 
\citet{Dalla_etal_2013}, drifts can significantly affect the energy of SEPs traveling scatter-free through the electromagnetic field of an ideal Parker wind.
Here we show that particles gain energy downstream of the quasi-perpendicular portion of a CME-driven shock, through the well-known 
energization of the gradient drift in the motional electric field. The mechanism is closely analogous to shock-drift acceleration, 
with the difference that, in our configuration, the dominant energy gain occurs in a finite region downstream of the shock rather than 
at the shock ramp itself. The novelty of the present work therefore does not reside in the acceleration mechanism itself but in its 
application and quantification in a 3D, time-dependent CME-spheromak configuration. We show that the energy gain can 
substantially exceed the cumulated losses due to collisions \citep{Ruffolo_1995,Giacalone_2002,Dalla_etal_2015,houeibib_2025} and drift motion 
in the unperturbed solar wind.
The paper is structured as follows. We present the MHD simulation and the GCA equations in Sect. \ref{sec:config}. We then describe the physical mechanism responsible 
for the acceleration of the particles in the CME driven shock in Sect. \ref{sec:energy}, discuss the role of scattering in 
Sect. \ref{subsec:energySpectra}, and summarize the results in Sect. \ref{sec:conclusion}.

\section{Model and numerical setup} \label{sec:config}
We used an MHD code to simulate a 3D, time-dependent, magnetized Parker-type wind perturbed by a CME. We then propagated energetic protons 
in the time-dependent electromagnetic field of the MHD simulation by numerically integrating the equations of motion in the guiding-center approximation.    

\begin{figure*}[!t]
\centering
\includegraphics[width=.8\textwidth]{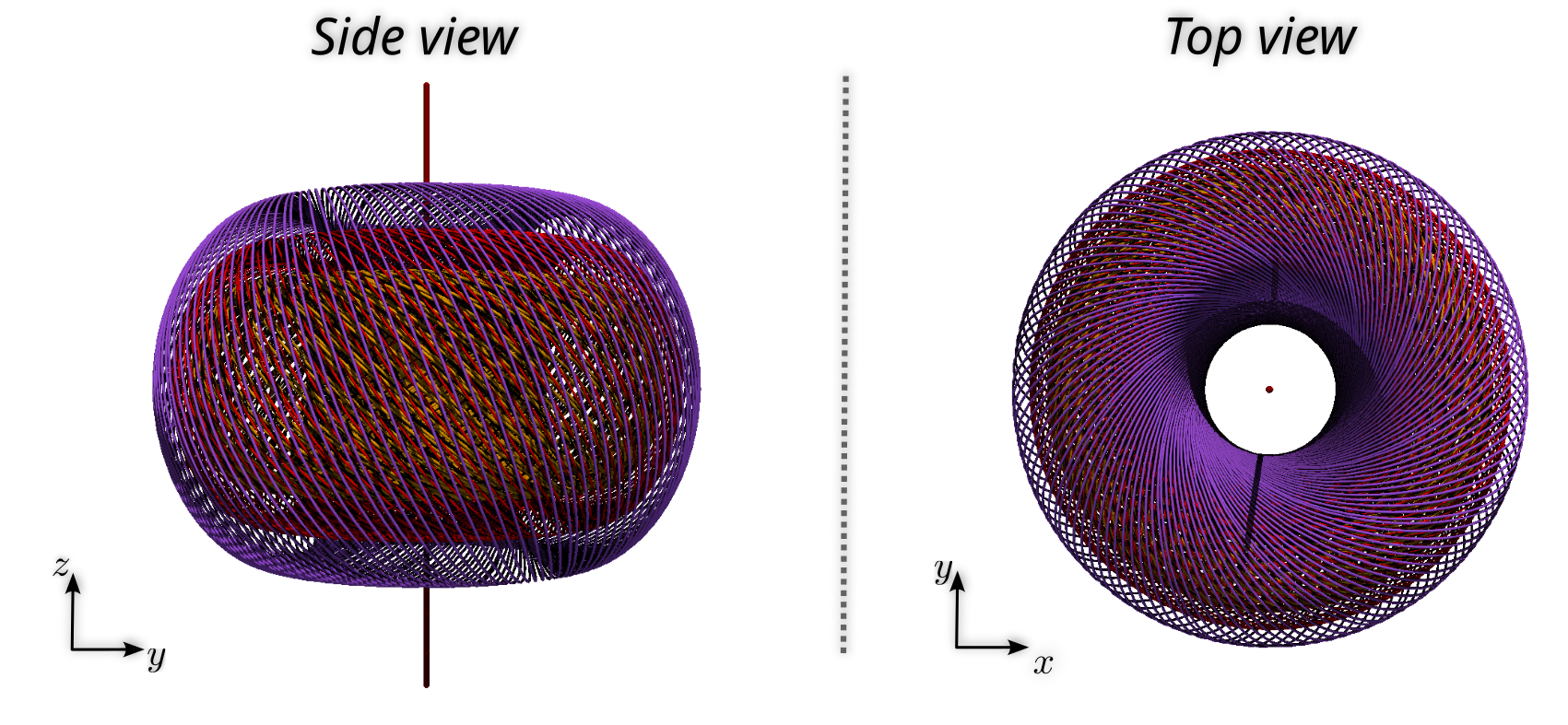}
        \caption{Top and side views of the magnetic field lines in a spheromak. Despite appearances, the magnetic field strength of the spheromak 
        increases monotonically toward its center, where it peaks at $2B_0/3$ (see equations (\ref{eq:spheromak_1}) and (\ref{eq:spheromak_2})). 
        At the spheromak boundary, $r^\prime=R_s$, only the poloidal component is nonzero, with $B_p^\prime = 0.2172\:B_0\sin(\theta^\prime)$.}
        \label{fig:spheromak}
\end{figure*}

\subsection{MHD model of the solar wind and CME} \label{subsec:mhd}
As in \cite{houeibib_2025}, we used the 3D MHD code MPI-AMRVAC (\citep{Keppens2023}) within the ideal-MHD approximation to simulate the solar 
wind and the CME. We treated the plasma as an ideal gas with a polytropic index of 5/3. 
\begin{figure}[!b]
\centering
\includegraphics[width=.4\textwidth]{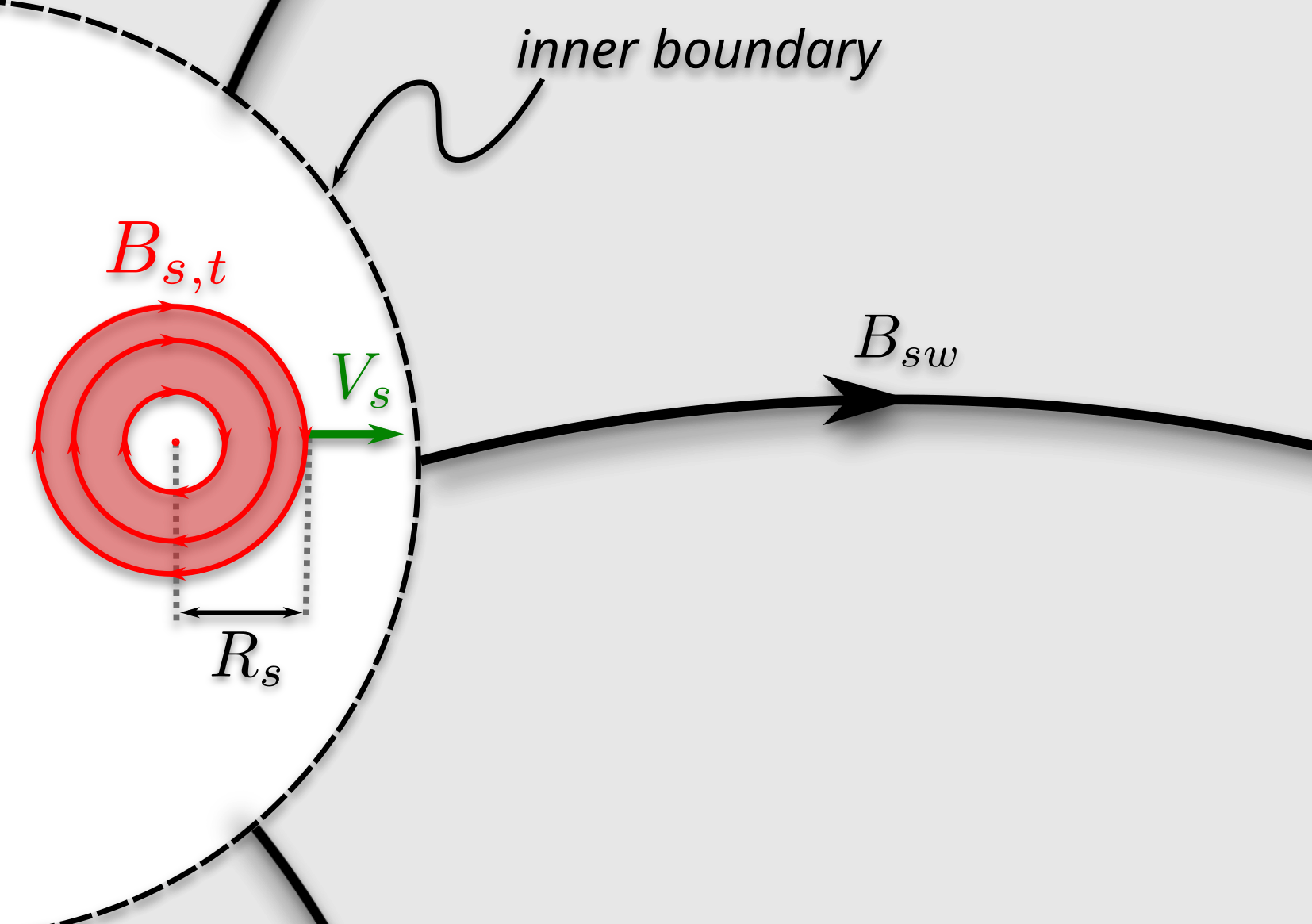}
\includegraphics[width=.4\textwidth]{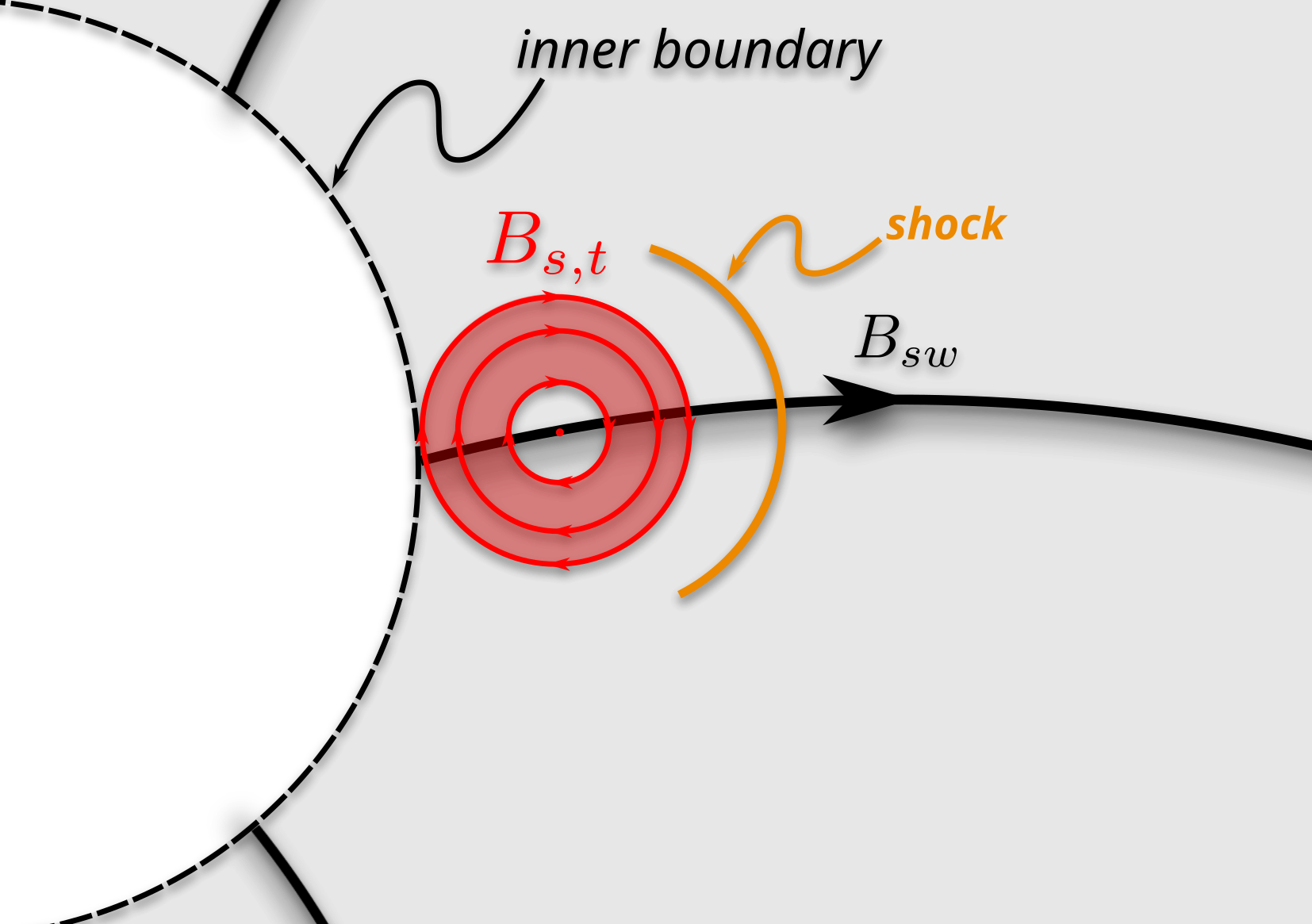}
\caption{Injection of the spheromak into the numerical domain. Top: Spheromak injection into the equatorial plane at constant radial speed $V_s$ into a preexisting stationary wind.}
\label{fig:configSpheromak}
\end{figure}
To convert mass density $\varrho$ and pressure $p$ to number density 
$n$ and temperature $T$ (which do not appear in the MHD equations), we assumed a fully ionized proton-electron plasma in which protons and electrons have the 
same temperature, $T_e=T_p=T$, and the same number density, $n_e=n_p=n/2$, which implies that $\varrho=\overline{m}n$, with  $\overline{m}\equiv(m_e+m_p)/2\simeq m_p/2$. 
The simulation domain is a Sun-centered spherical grid of size $[144 \times 48 \times 128]$ in $[r, \theta, \varphi]$, where $r$ is the radial coordinate
extending from $r$ = 0.139 AU to $r$ = 13.95 AU, $\theta$ is the polar angle, ranging from $0$ to $\pi$, and $\varphi$ is the azimuthal angle, ranging 
from $0$ to $2\pi$. The grid is uniform in $\theta$ and $\varphi$. To prevent  excessive longitudinal and latitudinal stretching of 
the cells, we adopted a stretching factor of 1.02 between adjacent cells in the direction of increasing $r$. 
For simplicity, we adopted a Parker-type wind and a positive magnetic monopole located at the Sun's center, so that there is no induced heliospheric current 
sheet, as in the case of a dipolar intrinsic field. At the Sun's surface, the monopole field strength is $2 \times 10^{-4} T$. 
We assumed that the inner boundary rotates rigidly with the Sun. Consequently, the tangential velocity components of the plasma at the inner boundary 
are prescribed as $\bm {\Omega} \times {\bm r}$, where $\bm \Omega = \Omega \hat{\mathbf{z}}$ with $\Omega = 2\pi/(30\,\mathrm{d})$, corresponding to the 
angular rotation speed of the Sun. At the inner boundary, we imposed a Neumann condition $\partial u_r / \partial r = 0$ to the radial component of the plasma 
speed $u_r$, a constant temperature $T=2\,\mathrm{MK}$, and a constant mass density $\varrho=1.0 \times 10^{-19}\,\mathrm{kg . \,m^{-3}}$. 
We also applied a Neumann condition to the radial magnetic-field component, $\partial B_r / \partial r = 0$, while the tangential components vanish at the inner boundary.
At the outer boundary, we imposed zero-gradient conditions, $\partial / \partial r = 0$, for all physical quantities. The inner boundary of the 
simulation domain lies beyond the sonic point, so the wind starts supersonic at the boundary. After several solar rotations, the simulation 
reaches a steady state. Table \ref{tab:swprop} lists the wind parameters at 1 AU in the equatorial plane of the simulation. The simulated wind is somewhat 
underdense but otherwise typical of the real solar wind at 1 AU \citep{Larrodera_Cid_2020,Salem_al_2023}.
\begin{table}[ht]
    \centering
    \caption{\label{tab:swprop} MHD simulation: Wind parameters in the equatorial plane at $r=1\:{\rm AU}$. Here, $p$ denotes the plasma thermal pressure, and  
    $\mu_0$ is the vacuum permeability.}
    \begin{tabular}{ll}
        \hline \hline \\
        Magnetic field strength $B$      & $ 4.96\: {\rm nT}$  \\
        Wind speed $u_{\rm sw}$          & $430\:{\rm km/s}$ \\
        Number density $n$ & $1.10\:{\rm cm^{-3}}$  \\
        Sound speed $c_{\rm s}=(\tfrac{5}{3}p/\varrho)^{1/2}$ & $48.62\:{\rm km/s}$ \\
        Alfvén speed $c_{\rm A}=(B^2/\mu_0\varrho)^{1/2}$     & $146.17\:{\rm km/s}$\\
        Plasma beta $\beta=p\:2\mu_0 /B^2$                                   & $0.13$\\
    \hline \hline
    \end{tabular}
\end{table}
\begin{figure*}[!t]
    \centering
    \includegraphics[width=\textwidth]{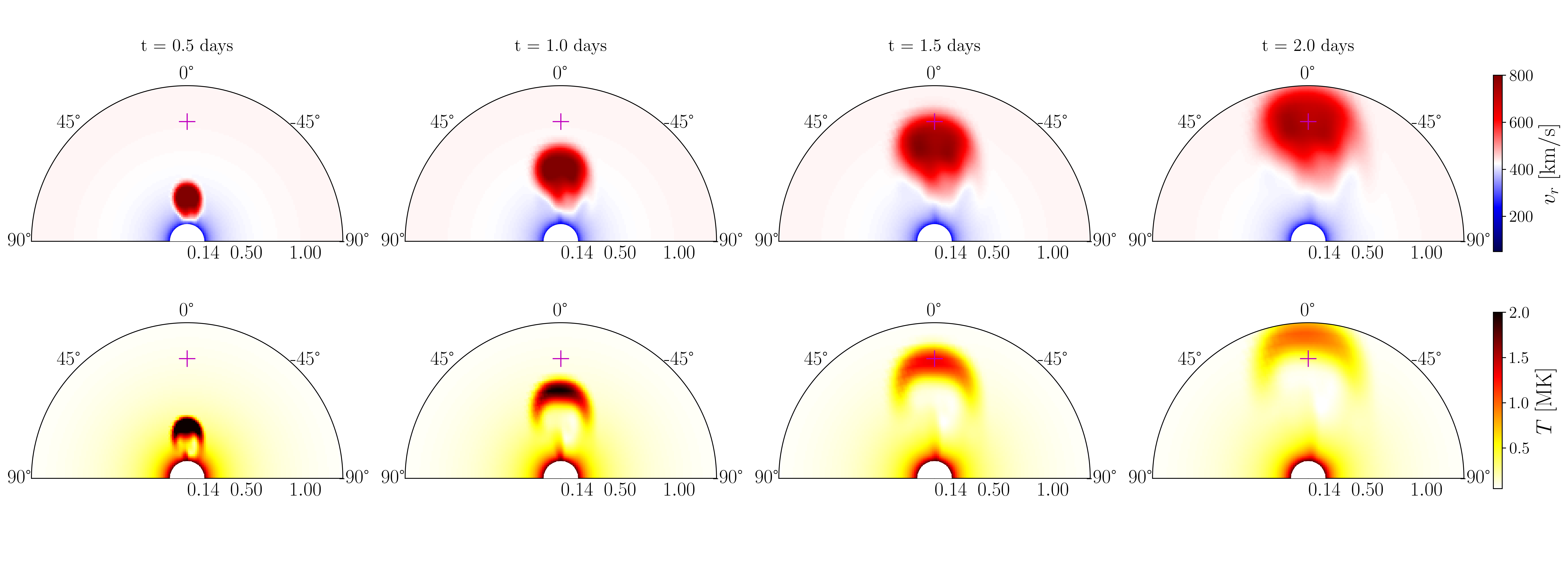}
    \caption{Equatorial cuts through the simulation domain as a function of time, where t=0 corresponds to the spheromak injection time. The color 
    maps show the radial fluid velocity (top) and temperature (bottom). The associated movie is available online.}
    \label{fig:slicesCME}  
\end{figure*}

Following previous authors \citep{Verbeke_2019,Kataoka_2009,Shiota_2016,Singh_2020,Koehn_2022} a CME is triggered by injecting a 
force-free spheromak-type magnetic structure through the inner domain boundary. 
The magnetic field of a spheromak is conveniently defined using a spherical coordinates
system ($r^\prime,\theta^\prime,\phi^\prime$), where the prime refers to 
the spheromak frame. In this frame, for $r^\prime\leq R_s$, the poloidal and toroidal components of the spheromak are given by    
\begin{eqnarray}
    \bm B_p^\prime &=& B_0\left[ \left( 2 \frac{j_1(\lambda r^{\prime})}{\lambda r^{\prime}} \cos \theta^{\prime} \right){\bm e_{r}^{\prime}} \right. \nonumber \\ 
    & & - \left. \left( j_0(\lambda r^{\prime})-\frac{j_1(\lambda r^{\prime})}{\lambda r^{\prime}} \right) \sin \theta^{\prime} {\bm e_{\theta}^{\prime}}\right], \label{eq:spheromak_1} \\ 
    \bm B_t^{\prime} &=& H B_0 j_1(\lambda r^{\prime}) \sin \theta{^{\prime}\bm e_{\phi}^{\prime}}. \label{eq:spheromak_2}  
\end{eqnarray}
For $r^\prime > R_s$, all components are set to zero. In equations (\ref{eq:spheromak_1}) and (\ref{eq:spheromak_2}), $B_0$ is a reference field strength, 
$j_{0,1}$ the first two spherical Bessel functions, and $\lambda\simeq 4.493409/R_s$, such that $j_1=0$ at the edge of the spheromak at $r^\prime=R_s$. 
At $r^\prime=R_s$, the toroidal field $B_t^{\prime}$ vanishes, while the poloidal field reduces to ${\bm B}_p^{\prime}=0.21723\:B_0\sin \theta^{\prime}{\bm e_{\theta}^{\prime}}$. The parameter
$H=\pm 1$ defines the handedness of the magnetic structure through the orientation of the toroidal field. 
In our simulation, we set $B_0 = 144$ nT and $R_s = 10.5 \: R_{\odot}$. The handedness of the spheromak is $H = +1$.
\begin{figure}[!htp]
    \centering
    \includegraphics[width=\columnwidth]{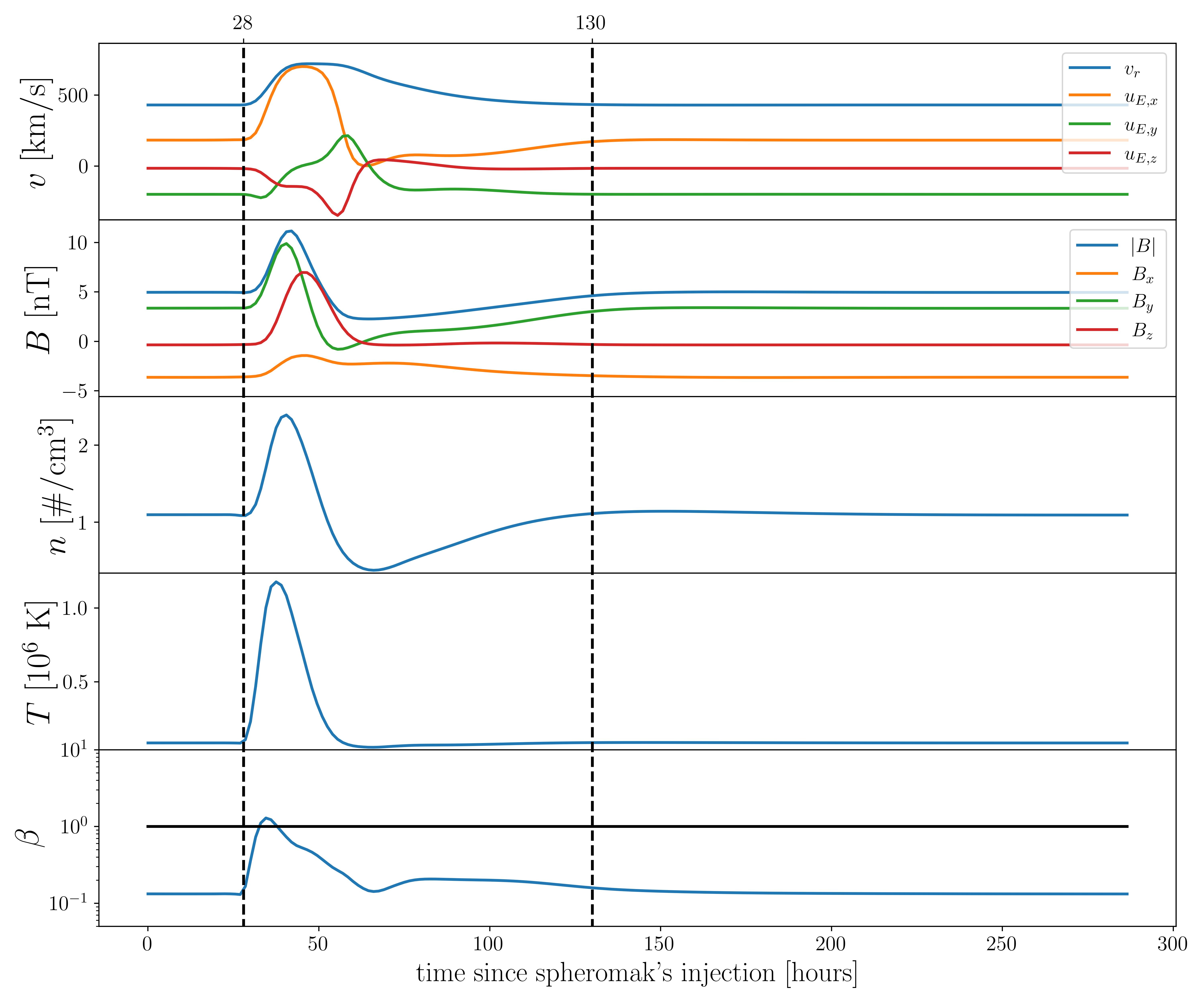}
    \caption{Variations of the plasma parameters induced by a CME at 1 AU. The vertical lines approximately delimit the time interval during which the local 
    plasma is considered to be perturbed by the CME.}
    \label{fig:CMEat1au}
\end{figure}
Figure \ref{fig:configSpheromak} shows that the spheromak is introduced into the simulation through its inner domain boundary in the equatorial plane 
(defined by the Sun's rotation) at a constant radial speed of $V_s = 1028$ km/s. We oriented the axis of the spheromak perpendicular to the 
equatorial plane. The mean plasma density and temperature in the spheromak are n = $65 \:{\rm cm^{-3}}$ and T = $0.6 \:{\rm {MK}}$, respectively. 
The equatorial cuts in Fig. \ref{fig:slicesCME} show temporal snapshots of the fluid temperature and radial velocity, which both trace the propagation of the 
shock front through its heating and acceleration of the fluid it encounters. The inner edge of the high-temperature arc also delimits the discontinuity separating 
the fluid heated by the shock from the magnetic driver of the CME, the spheromak ``remnant.'' The full time evolution of these equatorial cuts is available as an online movie.  

\begin{figure*}[!b]
    \centering
    \includegraphics[width=\textwidth]{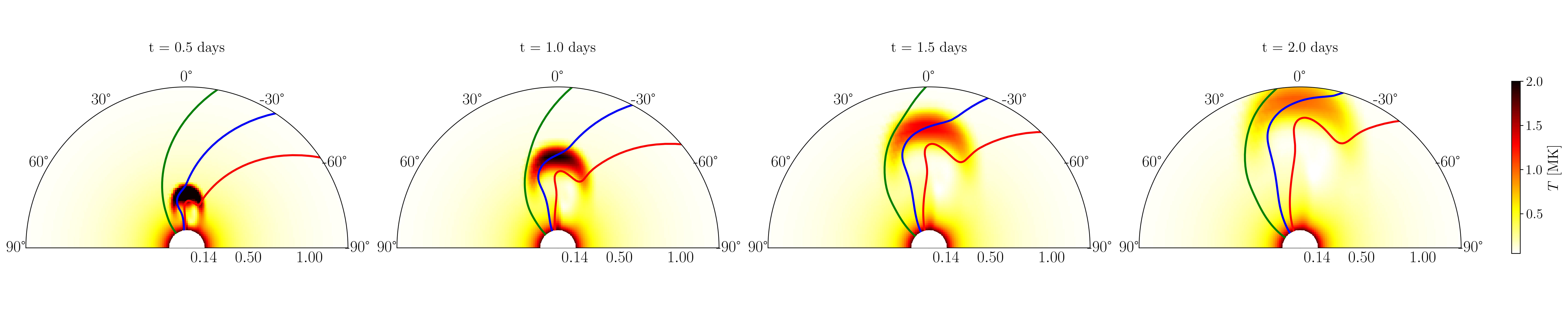}
    \caption{Time evolution of three selected magnetic field lines, denoted 0 (red line), 1 (blue line), and 2 (green line), from west to east. The associated movie is available online.}
    \label{fig:SourceLinesT}  
\end{figure*}

Figure \ref{fig:CMEat1au} shows the temporal profiles for various plasma parameters measured at a fixed position at 1 AU, represented by the cross in 
Figure \ref{fig:slicesCME}. The time sequence can be decomposed as follows:     
\begin{itemize}
    \item[--] Shock arrival: The shock reaches the observation point at $t \approx 28$ h after the spheromak's injection, when the density rises by a 
    factor of three and the magnetic field intensity increases from $\sim 5$ nT to $\sim 11$ nT. 
    \item[--] Sheath region: Between $t\approx 28\:{\rm h}$ and $t\approx 50\:{\rm h}$, the observation point is in a region of plasma compressed and heated by the shock. 
    \item[--] CME core (ejecta): Between $t\approx 50\:{\rm h}$ and $t\approx 130\:{\rm h}$, the observation point is in the ejecta, the spheromak remnant. 
    During the early phase of this period, the magnetic-field lines move eastward ($u_{E,y}>0$) instead of generally westward as in the unperturbed wind 
    (see also Figure \ref{fig:SourceLinesT}). For $t\gtrsim 130\:{\rm h}$, the profiles slowly recover pre-shock conditions.     
\end{itemize}

\subsection{Guiding-center equations}
In the following, we consider the motion of 5 GeV protons injected on three selected equatorial field lines in the CME perturbed solar wind described in 
Section \ref{sec:config}. Figure \ref{fig:SourceLinesT} shows  the evolution in time of the three field lines, numbered 0, 1, and 2. The corresponding time evolution is also available as an online movie. Unlike 
\cite{houeibib_2025} who considered steady fields, we used the guiding-center equations for time-varying $\bm E$ and $\bm B$ fields. In the limit 
$|\bm v_E|=|\bm{E}\times \bm{b}/B| \ll c$ , where $c$ is the speed of light, they can be written as follows: 
\begin{eqnarray}
    \label{eq:gca1}
    \frac{\mathrm{d}\bm R}{\mathrm{d}t} &=& v_{\|} {\bm b} + \bm v_E + \frac{\gamma m}{qB} \bm b \times \left[ \frac{v_{\|}}{\gamma}\frac{qE_{\|}}{mc^2}\bm v_E  \right. \nonumber \\ 
    &+& \left. \frac{\mu_{\rm B}}{\gamma^2 m} \left(\nabla B +  \frac{\bm v_E}{c^2} \partial_t B \right)  +  
    v_{\|} {\rm D}_t \bm b + {\rm D}_t \bm v_E
    \right] \\
    \label{eq:gca2}
    \frac{\mathrm{d} (\gamma v_{\|})}{\mathrm{d}t}  &=& \frac{q}{m}E_{\|} - 
    \frac{\mu_{\rm B}}{\gamma m}\bm b\cdot\nabla B + \gamma\bm v_E\cdot {\rm D}_t \bm b \\
    \label{eq:gca3}\frac{\mathrm{d}\mu_{\rm B}}{\mathrm{d}t}  &=& 0,  
\end{eqnarray}
where $ {\rm D}_t(*) = \partial_t(*) + (v_{\|}{\bm b} + \bm v_E) \cdot \nabla(*)$. We note that the factor $q/mc^2$ in (\ref{eq:gca1}) was mistakenly 
omitted in equation (1) of \cite{houeibib_2025}. In the above equations, the subscripts $\parallel$ and $\perp$ indicate projections parallel and 
perpendicular to $\bm{B}$.  The particle's guiding-center position is $\bm{R}$, its velocity is ${\bm v}$ , its Lorentz 
factor is $\gamma \equiv 1/\sqrt{1-v^2/c^2}$,  its magnetic moment is $\mu_B\equiv \tfrac{1}{2}m\gamma^2v_{\perp}^2/B $, $\bm{b}\equiv{\bm B}/B$, and $\bm{v}_{\rm E} \equiv \bm{E}\times \bm{b}/B$. 
We advanced the particles numerically in time using the third-order accurate predictor-corrector scheme used by \cite{houeibib_2025} (see also \cite{mignone_2023}), 
with all fields and derivatives of the fields on the right-hand side of the above GC equations computed within the MHD code and linearly interpolated from the MHD 
grid to the particle's position. 
The fields are also linearly interpolated in time between consecutive MHD snapshots. The integration time step $\Delta t$ is given by 
$\Delta t = {\rm min} \left( L_B,\lambda_{\parallel} \right) /c$, where $c$ is the speed of light, $L_B \equiv (|\nabla_{\|} B|/B)^{-1}$, and $\lambda_{\parallel}$ is
the scattering mean-free path (see Section \ref{subsec:energySpectra}). Although the structure of the GCA equations is much more complex than that of 
the full equations of motion, they offer a considerable advantage by eliminating the need for an integration time step smaller than the particle gyration 
period. This approach enables the integration of particle trajectories with higher precision over longer timescales at reduced computational costs.

\section{Time evolution of a proton in the CME field (no scattering)} \label{sec:energy}
\begin{figure}[!b]
    \centering
    \includegraphics[scale=0.78]{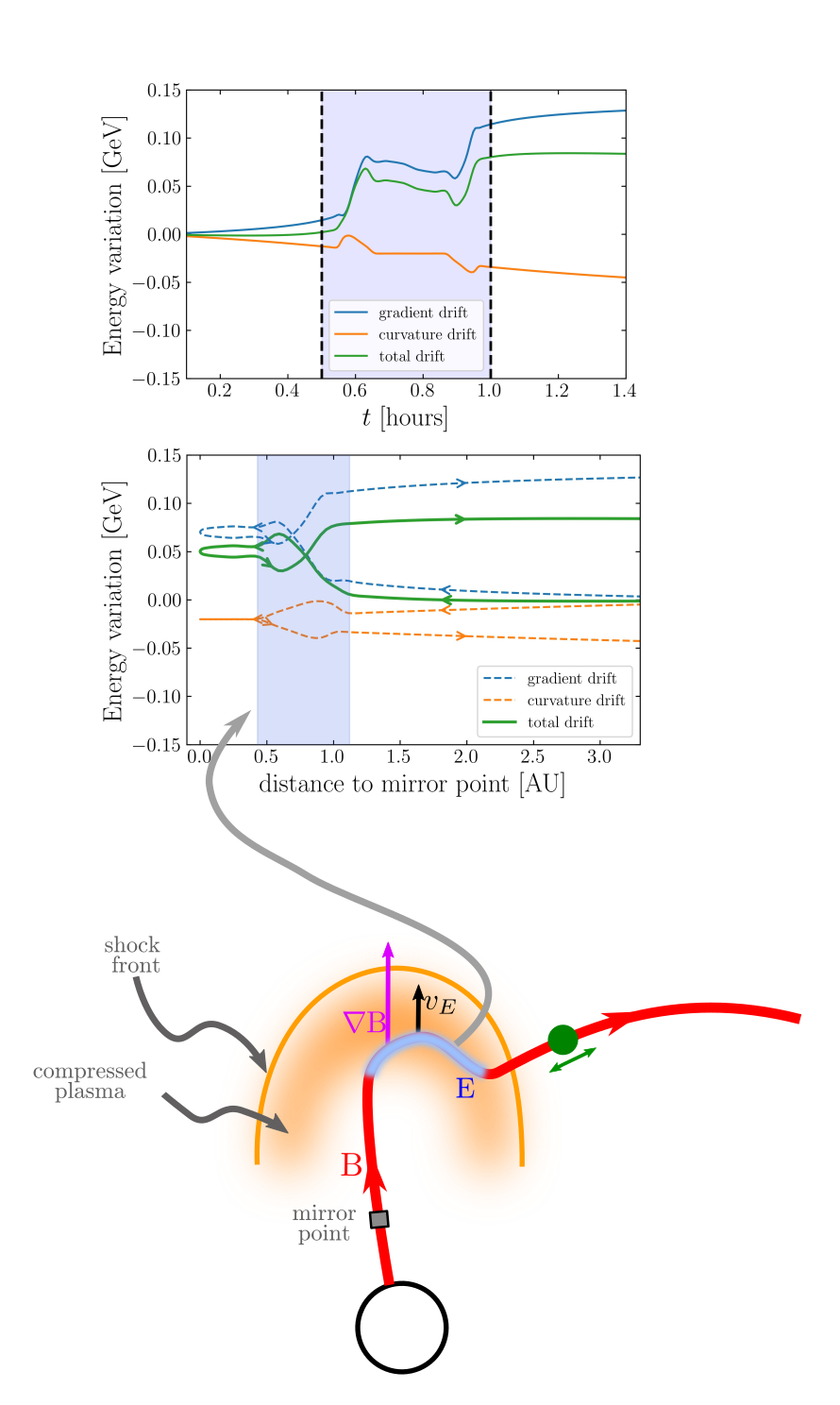}
    \caption{Top: Energy variation of a 5 GeV proton moving 
    throughout the CME-perturbed plasma along line 0 as a function of time. Here, 
    $t=0$ corresponds to the time at which the particle is injected into the system. The first 
    increment at $t\approx 0.6\,\mathrm{h}$ occurs as the particle moves toward the Sun, and
    the second at $t\approx 0.9\,\mathrm{h}$ as the particle moves away from the Sun after 
    reflection at its mirror point. The dotted vertical lines delimit the period during which the 
    particle is inside the CME-perturbed region. Middle: Energy variation as a 
    function of distance from the mirror point. The arrows indicate the direction of motion of 
    the particle. Bottom: Schematic representation of line 0.
    The particle (green dot) gains energy as it travels through the compressed plasma downstream of 
    the shock front, where $\nabla B \cdot \mathbf{v}_E > 0 $.}
    \label{fig:qvdE}
\end{figure}

\begin{figure}[!h]
    \centering
    \includegraphics[scale=0.65]{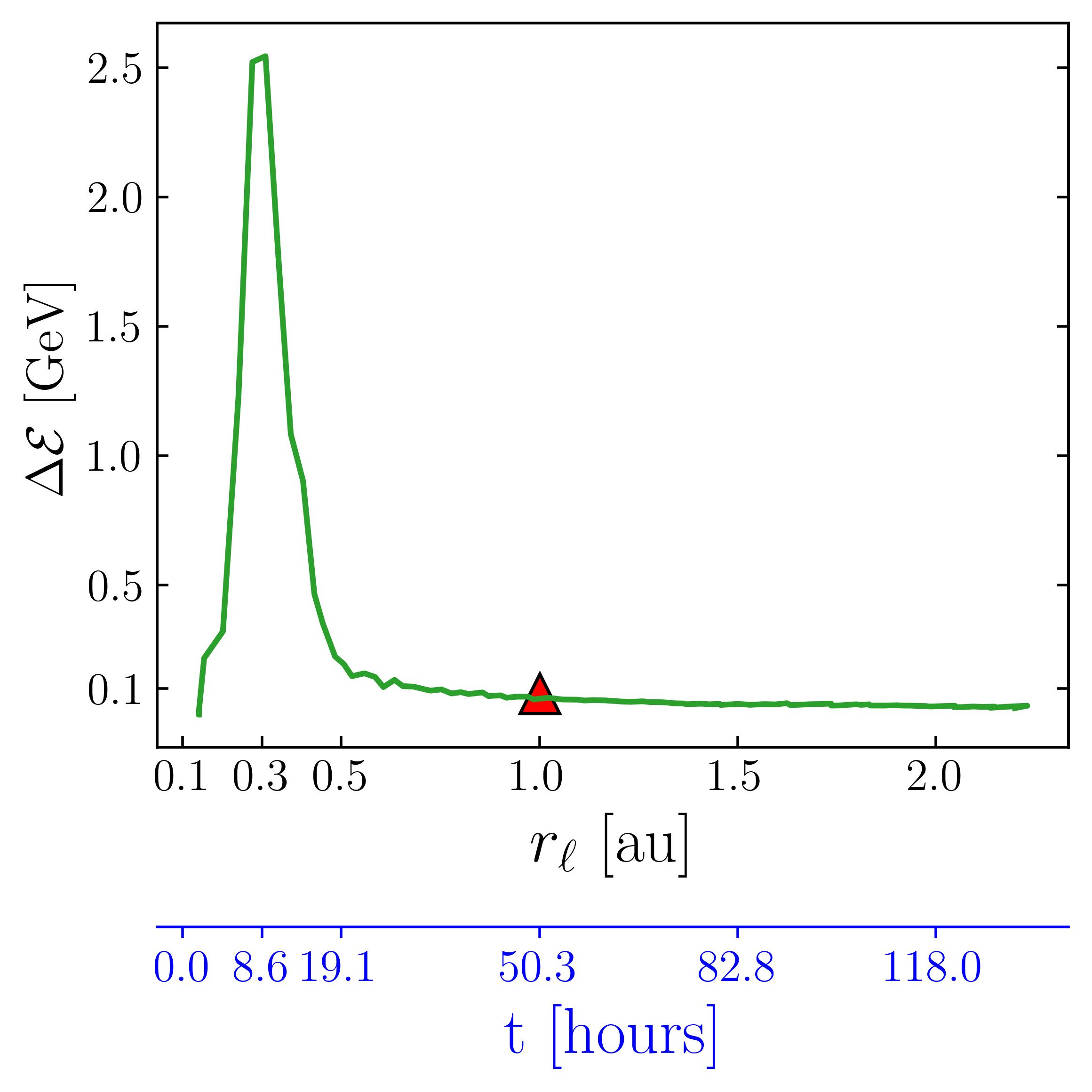}
    \caption{
    Energy gain estimated using Eq. (\ref{eq:ekgain}) for a fictitious proton moving along the portion of the line 0 located downstream of the quasi-perpendicular 
    shock of the CME at different phases of its expansion. The same 5 GeV proton as in Fig. \ref{fig:qvdE} is assumed. The time $t=0$ corresponds to the 
    injection of the CME through the inner boundary of the simulation.}
    \label{fig:ekLine}
\end{figure}

Let us first consider the propagation of a single 5 GeV proton injected onto the magnetic field line 0, some $\sim 1.5$ days after the spheromak injection 
(red line in the bottom-left panel of Fig. \ref{fig:SourceLinesT}). The particle is initially positioned at $ r = 3 \, \mathrm{AU}$ with a pitch-angle 
$\alpha = 170^\circ$ (traveling toward the Sun) so that it is mirror-reflected before reaching the inner boundary. The speed of a relativistic proton 
greatly exceeds that of the CME, so the fields can be assumed to be static during the proton's round trip, that is, $\partial /\partial t =0$ in (\ref{eq:gca2}). Because $E_{\|} \ll E_{\perp}$, as is generally verified in the solar wind, we conclude that the proton's kinetic energy,  $\mathcal{E}=(\gamma-1)mc^2$, evolves according to 
\begin{eqnarray}
    \label{eq:gcaEk}
    \frac{d\mathcal{E}}{dt} &\simeq  m\gamma v_\parallel^2\; 
    {\bm v}_{\rm E}\cdot({\bm b}\cdot\nabla){\bm b} \;+ \; m \gamma \frac{v_{\perp}^2}{2}\; 
    {\bm v}_{\rm E}\cdot \nabla {\rm ln} B \nonumber \\ 
    &\simeq  m\gamma v_\parallel^2\; 
    {\bm v}_{\rm E}\cdot({\bm b}\cdot\nabla){\bm b} \;+ \; \frac{\mu_B}{\gamma}\; 
    {\bm v}_{\rm E}\cdot \nabla B . 
\end{eqnarray}
In the above equation, the terms on the right-hand side correspond to energy gains due to the curvature and gradient drifts, respectively. 
These terms are contained in Eq. \ref{eq:gca1} and can be written as
\begin{eqnarray}
    {\bm v}_{\rm curv} = \frac{\gamma m v_\parallel^2}{qB} {\bm b}\times({\bm b}\cdot\nabla){\bm b}, \qquad
    {\bm v}_{\nabla B} = \frac{\gamma m v_\perp^2}{2qB} {\bm b}\times\nabla\ln B. \nonumber
\end{eqnarray}
Because the energy variation $d\mathcal{E}/dt$ associated with each term can be written as $q{\bm v_d}\cdot{\bm E}$, where ${\bm v_d}$ is a generic drift velocity,
we plotted their relative contributions as a function of time (top panel of Fig. \ref{fig:qvdE}) and distance from the mirror point 
(middle panel of Fig. \ref{fig:qvdE}). The acceleration is clearly attributable to the gradient-drift term.
We denote the time interval of strong acceleration by $\Delta t_c\approx 0.05\:{\rm h}$. This interval comprises two stages, at $t\approx 0.6\:{\rm h}$ and 
$t\approx 0.9\:{\rm h}$ (see the top panel of Figure \ref{fig:qvdE}). Both acceleration phases occur while the particle is located 
downstream of the shock, where ${\bm v}_{\rm E}\cdot \nabla B > 0$. The first acceleration step takes place as the particle moves inward (toward the Sun), and
the second as it moves outward after reflection at its mirror point. 
The energy gain can therefore be written as    
\begin{eqnarray}
    \label{eq:ekgain}
    \Delta \mathcal{E} \simeq \frac{\mu_B}{\gamma} \Delta t_c {\bm v_E} \cdot \nabla B, 
\end{eqnarray}
or, in terms of relative variation, as   
\begin{eqnarray}
    \label{eq:ekgainodg}
    \frac{\Delta \mathcal{E}}{\mathcal{E}} \simeq  \frac{1}{2}\:\frac{\gamma^2}{\gamma(\gamma-1)}\:\frac{v_\perp^2}{c^2}\: \Delta t_c 
    {\bm v_E} \cdot \frac{\nabla B}{B}\propto \Delta t_c 
    {\bm v_E} \cdot \frac{\nabla B}{B}.
\end{eqnarray}
Equations (\ref{eq:ekgain}) and (\ref{eq:ekgainodg}) show that the energy gain is proportional to $\Delta t_c$, that is, proportional to the time the particle spends in the region where ${\bm v_E} \cdot \nabla B>0$. 
Fig. \ref{fig:ekLine} traces the energy gain of a fictitious 5 GeV proton (with the same $\mu_B$ as in Fig. \ref{fig:qvdE}), 
moving along field line 0 at different phases of the CME's expansion. 
\begin{figure}[!h]
    \centering
    \includegraphics[scale=0.45]{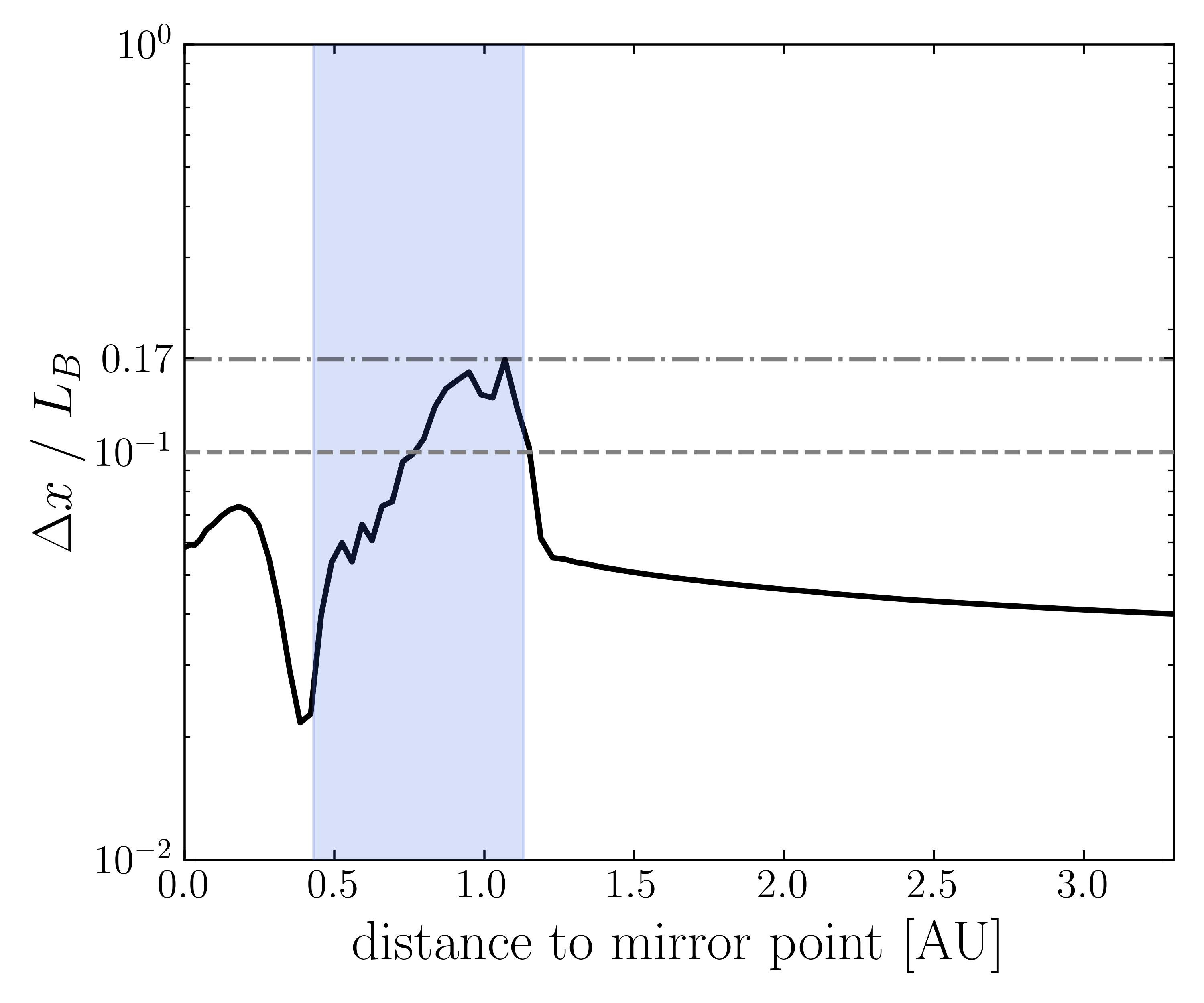}
    \caption{Ratio of the local grid spacing $\Delta x$ to the magnetic gradient scale length $L_B$ along 
    the trajectory of the proton in Fig. \ref{fig:qvdE}, as a function of the distance to the mirror point. 
    The shaded area indicates the acceleration region shown in the middle panel of Fig. \ref{fig:qvdE}.}
    \label{fig:resolution}
\end{figure}
Clearly, efficiency peaks at the time when the CME reaches $r \simeq 0.3$ AU, when the product of the three terms $\Delta t_c {\bm v_E}\cdot\nabla B$ in equation 
(\ref{eq:ekgain}) is at its maximum. The red diamond represents the expected gain for a particle that encounters the CME at 1 AU, corresponding to the gain of the particle in Fig. \ref{fig:qvdE}. 

Since energization is proportional to the magnetic-field gradient through $\bm{v}_E \cdot \nabla B$, we investigated whether the gradient scale length $L_B \equiv (|\nabla_\parallel B|/B)^{-1}$ in the acceleration region 
is determined by CME-driven compression or by numerical diffusion. Fig.~\ref{fig:resolution} shows the ratio 
$\Delta x / L_B$ of the local grid spacing to $L_B$ along the trajectory of the particle shown in Fig. \ref{fig:qvdE}. 
The ratio remains below 0.17 everywhere; that is, $L_B$ is resolved by at least six grid cells, even at the 
strongest gradients encountered by the particle. The acceleration region itself extends over $\sim 0.7$~AU, 
corresponding to several tens of grid cells and extends well beyond the few cells over which the shock is numerically broadened. 
The gradients responsible for the energization are therefore not controlled by numerical diffusion.

\section{Protons in the CME field (with scattering)} \label{subsec:energySpectra}
In Section \ref{sec:energy}, we did not consider scattering in velocity space due to the particle's interaction with small-scale plasma turbulence 
not retained in the large-scale MHD simulation. In the absence of scattering, the injected particles pass through the acceleration region at most twice: 
first as they stream toward the Sun, and second time as they stream away from the Sun after reflection at their mirror point. In the presence of 
scattering, multiple crossings become possible. We therefore applied pitch-angle scattering to the particles with a parallel mean free path $\lambda_{\|}$. We considered two 
different values, 0.1 AU and 0.5 AU, which covered the generally accepted range for a wide range of energies at 1 AU \citep{palmer_transport_1982,bieber_proton_1994}. 
Particles undergo hard-sphere scattering by scattering centers at rest in the solar-wind frame (see Sect. 2.3 in \cite{houeibib_2025} for details).
The post-collision pitch-angle cosine is drawn as $\mu = \pm \sqrt{a}$, where $a$ is uniformly distributed over the range $[0,1]$ 
(see Sect. 2.3 in \cite{houeibib_2025} for a discussion of this point).
To accumulate statistics, we considered $10^4$ monoenergetic 5 GeV protons that were impulsively injected toward the Sun at $t=0$ h with a pitch angle of $\alpha = 180 ^\circ$. 
As in \citet{houeibib_2025}, particles that reach the boundaries at r=0.14 AU and r=3 AU are immediately reinjected at r = 3 AU along the same field line and with the same 
initial conditions, implying a constant number of particles in the domain.    
\begin{figure}[!htp]
    \includegraphics[scale=0.65]{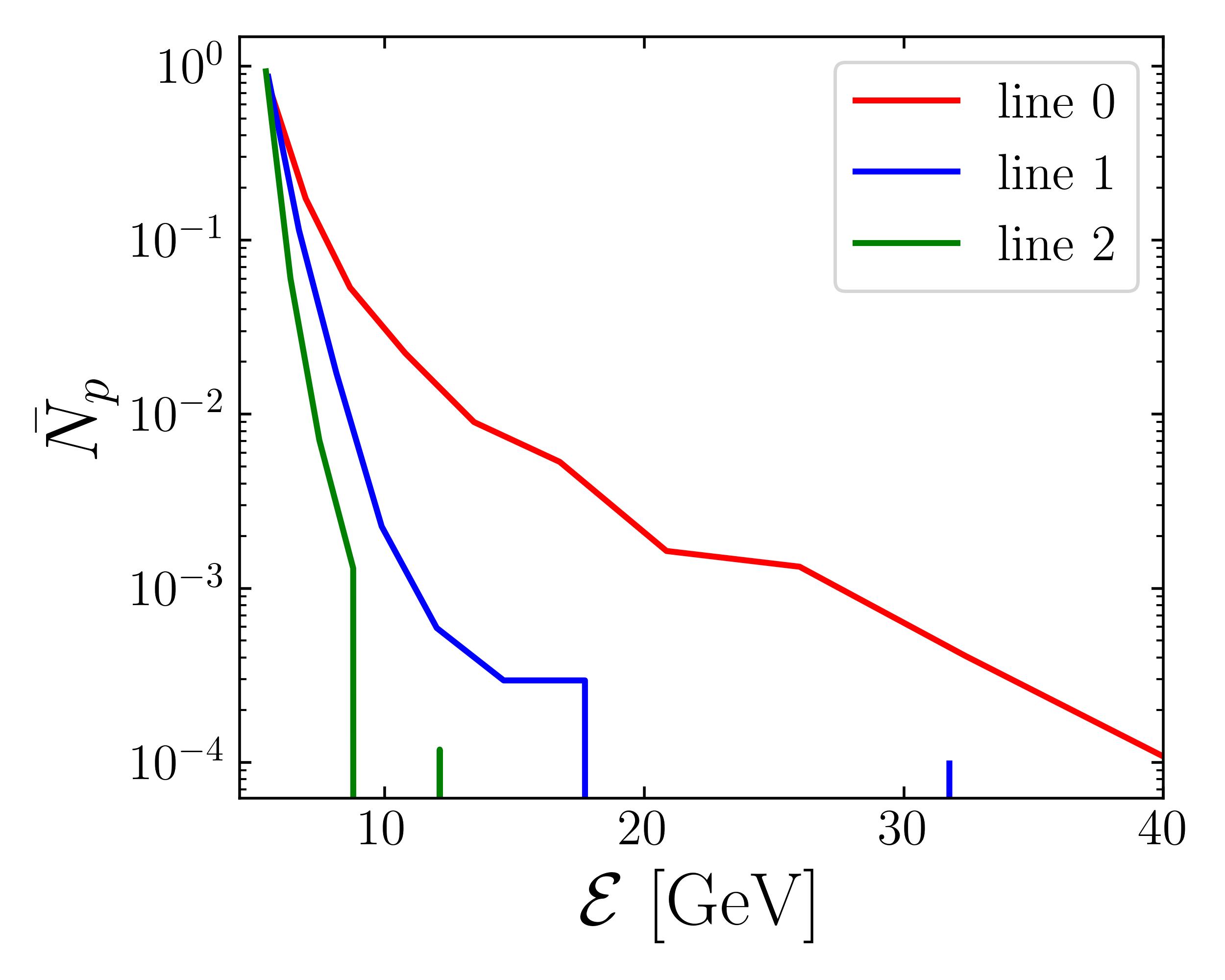}
    \caption{Cumulative energy spectra measured at 1 AU over a time interval of $\sim 4$ days for 5 GeV protons injected along the three different field lines and
    subject to scattering with a mean free path $\lambda_{\|} = 0.5$ AU.}
    \label{fig:E_spectrum}
\end{figure} 
\begin{figure}[!htp]
    \includegraphics[scale=0.65]{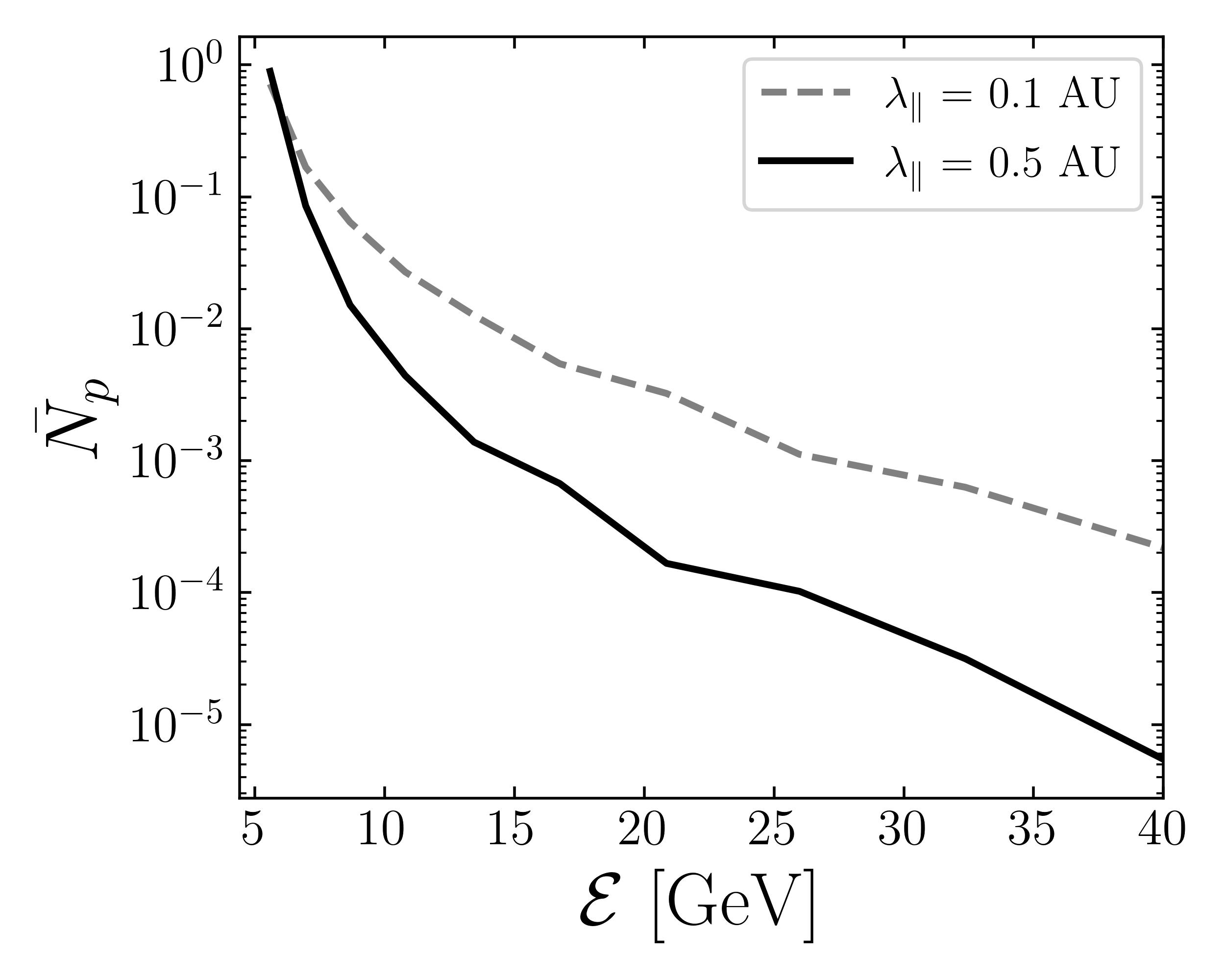}
    \caption{Cumulative energy spectra of particles at 1 AU on line 0 for two values of $\lambda_{\|}$ over a time interval of $\sim 4$ days. The difference 
    between the two curves is compatible with a $\lambda_{\|}^{-3/2}$ scaling factor. 
    }
    \label{fig:E_spectrum_diff}
\end{figure}
A particle subject to scattering may repeatedly pass through the acceleration region as it bounces back and forth between scattering centers or between mirror 
points and scattering centers. On some privileged magnetic field lines, particles gain energy at each crossing; consequently, they can reach much larger energies
than in the no-scattering case discussed in Section \ref{sec:energy}, provided the acceleration due to ${\bm v_E} \cdot \nabla B>0$ is sufficiently strong and long-lasting.
To illustrate the role of scattering, Fig. \ref{fig:E_spectrum} shows the four-day cumulative energy distributions for protons injected along field lines 0, 1, and 2 for $\lambda_{\|} = 0.5$ AU. We obtained the distributions by measuring the particles' energies as they cross the spherical 
shell at a radius of 1 AU. The same particle can contribute multiple times to the distribution if it crosses the 1 AU shell more than once.
We stress that these cumulative spectra are diagnostics of the adopted numerical experiment rather than direct predictions of an observable spectrum: the reinjection 
procedure at $r = 3$~AU maintains a driven, recirculating particle population and thereby controls the particle residence time and the number of possible 
passages through the acceleration region.
The figure shows that the acceleration is highest on line 0 and lowest on line 2. This is not surprising, as the three field lines are affected very differently by the CME, with line 0 having the longest portion in the region of strong magnetic compression (see Figure \ref{fig:SourceLinesT}).   
We expect that for reasonable values of $\lambda_{\|}$, the spectra harden with decreasing $\lambda_{\|}$. Indeed, reducing $\lambda_{\|}$ increases the particles' 
residence time in the acceleration region (also see \cite{vainio_2000}). The effect is illustrated in Fig. \ref{fig:E_spectrum_diff}, which shows the cumulative 
spectra on line 0 for $\lambda_{\|}=0.5$ AU and 0.1 AU, respectively. A semi-quantitative explanation can be given by noting that, in a 
1D diffusion experiment in which $N$ particles moving at speed c are instantaneously injected at $x=0$, the number density at any given point in space decreases 
asymptotically with time $t$ as $(\pi \lambda_{\|}ct)^{-1/2}$. The number of collisions experienced by a particle increases as $N_c=ct/\lambda_{\|}$,  and the energy gain $\Delta \mathcal{E}$ for a particle is approximately proportional to the number of collisions , such that $\Delta \mathcal{E}=N_c \delta \mathcal{E}$, where $\delta \mathcal{E}$ 
is the energy gain per passage in the acceleration region. Therefore, the number of particles that gain a given energy 
$\delta \mathcal{E}$ scales as $(ct/\lambda_{\|}^3)^{1/2} \propto  \lambda_{\|}^{-3/2}$. Consequently, reducing the mean free path increases the number of particles 
that gain a given energy $\Delta \mathcal{E}$. In our case, the number of particles accelerated to a given energy is expected to be larger by a factor of
$(0.5/0.1)^{3/2} \simeq 11.2$ for $\lambda_{\|}=0.1\:{\rm AU}$ than for  $\lambda_{\|}=0.5\:{\rm AU}$. This factor effectively 
separates the two curves in Fig. \ref{fig:E_spectrum_diff}.

\section{Conclusions\label{sec:conclusion}}

We computed the guiding-center trajectories of relativistic 5 GeV protons in the field of an MHD simulation of a Parker-type solar wind perturbed by an equatorial CME. We triggered the CME by gradually inserting a spheromak through the inner boundary of the simulation domain at $r=0.139$ AU. We injected the protons sunward from a heliocentric 
distance of 3 AU along three equatorial magnetic field lines crossing the CME at different heliographic longitudes. We summarize our main findings below. 
\begin{enumerate}
    \item Particles gain energy as they travel through the compressed plasma downstream of the CME-driven shock. The acceleration results from the work done by the motional 
    electric field on the gradient drift, which is positive ($q\bm{v}_{\nabla B} \cdot \bm{E} > 0$, or equivalently $\bm v_E \cdot \nabla B > 0$) when the particle 
    is located downstream of the shock.
    \item A peak energy gain of up to $\sim$ 50 \% for a single passage through the acceleration region occurs when the CME shock front reaches 0.3 AU. 
    \item In the case of pitch-angle scattering, particles may pass through the acceleration region multiple times and reach substantially higher energies. For example, 
    assuming hard-sphere scattering and a mean free path $\lambda_{\|}=0.1\:{\rm AU}$, approximately $0.1\%$ of the injected particles increase their energy by a factor of six over four days.
    These figures are specific to our numerical experiment, in which particles are continuously reinjected at $r = 3$~AU, and should not be interpreted as predictions of observable spectra.
    \item The acceleration efficiency strongly depends on where the particle encounters the shock. For the three equatorial field lines considered, the efficiency 
    increases westward (see Figures \ref{fig:SourceLinesT} and \ref{fig:E_spectrum}).         
    \item The energy distributions harden for decreasing $\lambda_{\|}$ as the number of times particles pass through the acceleration region increases. 
    The number of particles reaching a given energy scales as $\lambda_{\|}^{-3/2}$, consistent with a heuristic 1D diffusion estimate.
\end{enumerate}

\begin{acknowledgements}
    This work has been financially supported by the PLAS@PAR project and by the National Institute of Sciences of the Universe (INSU). 
    AH is supported by the CNES (Centre National d'Études Spatiales).
\end{acknowledgements}

\bibliographystyle{aa}
\bibliography{aa}

@article{houeibib_2025,
  author  = {Houeibib, A. and Pantellini, F. and Griton, L.},
  title   = {Dynamics of energetic electrons scattered in the solar wind -- Magnetohydrodynamics and test-particle simulations},
  journal = {Astronomy \& Astrophysics},
  year    = {2025},
  month   = feb,
  volume  = {694},
  pages   = {A211},
  doi     = {10.1051/0004-6361/202451436},
}

@article{bieber_proton_1994,
  author  = {Bieber, John W. and Matthaeus, William H. and Smith, Charles W. and Wanner, Wolfgang and Kallenrode, May-Britt and Wibberenz, Gerd},
  title   = {Proton and Electron Mean Free Paths: The {Palmer} Consensus Revisited},
  journal = {The Astrophysical Journal},
  year    = {1994},
  month   = jan,
  volume  = {420},
  pages   = {294},
  doi     = {10.1086/173559},
}

@article{palmer_transport_1982,
  author  = {Palmer, I. D.},
  title   = {Transport coefficients of low-energy cosmic rays in interplanetary space},
  journal = {Reviews of Geophysics},
  year    = {1982},
  volume  = {20},
  number  = {2},
  pages   = {335--351},
  doi     = {10.1029/RG020i002p00335},
}

@article{Larrodera_Cid_2020,
  author  = {Larrodera, C. and Cid, C.},
  title   = {Bimodal distribution of the solar wind at 1 {AU}},
  journal = {Astronomy \& Astrophysics},
  year    = {2020},
  month   = mar,
  volume  = {635},
  pages   = {A44},
  doi     = {10.1051/0004-6361/201937307},
}

@article{Salem_al_2023,
  author  = {Salem, Chadi S. and Pulupa, Marc and Bale, Stuart D. and Verscharen, Daniel},
  title   = {Precision electron measurements in the solar wind at 1 au from {NASA}'s {Wind} spacecraft},
  journal = {Astronomy \& Astrophysics},
  year    = {2023},
  month   = jul,
  volume  = {675},
  pages   = {A162},
  doi     = {10.1051/0004-6361/202141816},
}

@article{mignone_2023,
  author  = {Mignone, A. and Haudemand, H. and Puzzoni, E.},
  title   = {A guiding center implementation for relativistic particle dynamics in the {PLUTO} code},
  journal = {Computer Physics Communications},
  year    = {2023},
  volume  = {285},
  pages   = {108625},
  doi     = {10.1016/j.cpc.2022.108625},
}

@article{Kilpua_2017,
  author  = {Kilpua, Emilia and Koskinen, Hannu E. J. and Pulkkinen, Tuija I.},
  title   = {Coronal mass ejections and their sheath regions in interplanetary space},
  journal = {Living Reviews in Solar Physics},
  year    = {2017},
  month   = dec,
  volume  = {14},
  number  = {1},
  pages   = {5},
  doi     = {10.1007/s41116-017-0009-6},
}

@article{Verbeke_2019,
  author  = {Verbeke, C. and Pomoell, J. and Poedts, S.},
  title   = {The evolution of coronal mass ejections in the inner heliosphere: Implementing the spheromak model with {EUHFORIA}},
  journal = {Astronomy \& Astrophysics},
  year    = {2019},
  month   = jul,
  volume  = {627},
  pages   = {A111},
  doi     = {10.1051/0004-6361/201834702},
}

@article{Kataoka_2009,
  author  = {Kataoka, R. and Ebisuzaki, T. and Kusano, K. and Shiota, D. and Inoue, S. and Yamamoto, T. T. and Tokumaru, M.},
  title   = {Three-dimensional {MHD} modeling of the solar wind structures associated with 13 December 2006 coronal mass ejection},
  journal = {Journal of Geophysical Research: Space Physics},
  year    = {2009},
  month   = oct,
  volume  = {114},
  number  = {A10},
  pages   = {A10102},
  doi     = {10.1029/2009JA014167},
}

@article{Shiota_2016,
  author  = {Shiota, D. and Kataoka, R.},
  title   = {Magnetohydrodynamic simulation of interplanetary propagation of multiple coronal mass ejections with internal magnetic flux rope ({SUSANOO-CME})},
  journal = {Space Weather},
  year    = {2016},
  month   = feb,
  volume  = {14},
  number  = {2},
  pages   = {56--75},
  doi     = {10.1002/2015SW001308},
}

@article{Singh_2020,
  author  = {Singh, Talwinder and Yalim, Mehmet S. and Pogorelov, Nikolai V. and Gopalswamy, Nat},
  title   = {A Modified Spheromak Model Suitable for Coronal Mass Ejection Simulations},
  journal = {The Astrophysical Journal},
  year    = {2020},
  month   = may,
  volume  = {894},
  number  = {1},
  pages   = {49},
  doi     = {10.3847/1538-4357/ab845f},
}

@article{Koehn_2022,
  author  = {Koehn, G. J. and Desai, R. T. and Davies, E. E. and Forsyth, R. J. and Eastwood, J. P. and Poedts, S.},
  title   = {Successive Interacting Coronal Mass Ejections: How to Create a Perfect Storm},
  journal = {The Astrophysical Journal},
  year    = {2022},
  month   = dec,
  volume  = {941},
  number  = {2},
  pages   = {139},
  doi     = {10.3847/1538-4357/aca28c},
}

@article{Keppens2023,
  author  = {Keppens, R. and Popescu Braileanu, B. and Zhou, Y. and Ruan, W. and Xia, C. and Guo, Y. and Claes, N. and Bacchini, F.},
  title   = {{MPI-AMRVAC} 3.0: Updates to an open-source simulation framework},
  journal = {Astronomy \& Astrophysics},
  year    = {2023},
  volume  = {673},
  pages   = {A66},
  doi     = {10.1051/0004-6361/202245359},
}

@article{Ruffolo_1995,
  author  = {Ruffolo, D.},
  title   = {Effect of Adiabatic Deceleration on the Focused Transport of Solar Cosmic Rays},
  journal = {The Astrophysical Journal},
  year    = {1995},
  month   = apr,
  volume  = {442},
  pages   = {861},
  doi     = {10.1086/175489},
}

@article{Reames_1997,
  author  = {Reames, Donald V.},
  title   = {Energetic Particles and the Structure of Coronal Mass Ejections},
  journal = {Geophysical Monograph Series},
  year    = {1997},
  month   = jan,
  volume  = {99},
  pages   = {217--226},
  doi     = {10.1029/GM099p0217},
}

@article{Klein_2017,
  author  = {Klein, Karl-Ludwig and Dalla, Silvia},
  title   = {Acceleration and Propagation of Solar Energetic Particles},
  journal = {Space Science Reviews},
  year    = {2017},
  month   = nov,
  volume  = {212},
  number  = {3-4},
  pages   = {1107--1136},
  doi     = {10.1007/s11214-017-0382-4},
}

@article{Desai_2016,
  author  = {Desai, Mihir and Giacalone, Joe},
  title   = {Large gradual solar energetic particle events},
  journal = {Living Reviews in Solar Physics},
  year    = {2016},
  month   = dec,
  volume  = {13},
  number  = {1},
  pages   = {3},
  doi     = {10.1007/s41116-016-0002-5},
}

@article{Dalla_etal_2013,
  author  = {Dalla, S. and Marsh, M. S. and Kelly, J. and Laitinen, T.},
  title   = {Solar energetic particle drifts in the {Parker} spiral},
  journal = {Journal of Geophysical Research: Space Physics},
  year    = {2013},
  month   = oct,
  volume  = {118},
  number  = {10},
  pages   = {5979--5985},
  doi     = {10.1002/jgra.50589},
}

@article{Dalla_etal_2015,
  author  = {Dalla, S. and Marsh, M. S. and Laitinen, T.},
  title   = {Drift-induced Deceleration of Solar Energetic Particles},
  journal = {The Astrophysical Journal},
  year    = {2015},
  month   = jul,
  volume  = {808},
  number  = {1},
  pages   = {62},
  doi     = {10.1088/0004-637X/808/1/62},
}

@article{Marsh_etal_2013,
  author  = {Marsh, M. S. and Dalla, S. and Kelly, J. and Laitinen, T.},
  title   = {Drift-induced Perpendicular Transport of Solar Energetic Particles},
  journal = {The Astrophysical Journal},
  year    = {2013},
  month   = aug,
  volume  = {774},
  number  = {1},
  pages   = {4},
  doi     = {10.1088/0004-637X/774/1/4},
}

@article{Vainio_2000,
  author  = {Vainio, R. and Kocharov, L. and Laitinen, T.},
  title   = {Interplanetary and Interacting Protons Accelerated in a Parallel Shock Wave},
  journal = {The Astrophysical Journal},
  year    = {2000},
  month   = jan,
  volume  = {528},
  number  = {2},
  pages   = {1015--1025},
  doi     = {10.1086/308202},
}

@article{Giacalone_2002,
  author  = {Giacalone, J. and Jokipii, J. R. and K{\'o}ta, J.},
  title   = {Particle Acceleration in Solar Wind Compression Regions},
  journal = {The Astrophysical Journal},
  year    = {2002},
  month   = jul,
  volume  = {573},
  number  = {2},
  pages   = {845--850},
  doi     = {10.1086/340660},
}

@article{leRoux2009,
  author  = {{le Roux}, J. A. and Webb, G. M.},
  title   = {Time-dependent acceleration of interstellar pickup ions at the heliospheric termination shock using a focused transport approach},
  journal = {The Astrophysical Journal},
  year    = {2009},
  volume  = {693},
  number  = {1},
  pages   = {534--551},
  doi     = {10.1088/0004-637X/693/1/534},
}

@article{leRoux2012,
  author  = {{le Roux}, J. A. and Webb, G. M.},
  title   = {A Focused Transport Approach to the Time-Dependent Shock Acceleration of Solar Energetic Particles at a Fast Traveling Shock},
  journal = {The Astrophysical Journal},
  year    = {2012},
  month   = feb,
  volume  = {746},
  number  = {1},
  pages   = {104},
  doi     = {10.1088/0004-637X/746/1/104},
}

@article{vandenBerg2020,
  author  = {{van den Berg}, J. and Strauss, D. T. and Effenberger, F.},
  title   = {A Primer on Focused Solar Energetic Particle Transport. Basic Physics and Recent Modelling Results},
  journal = {Space Science Reviews},
  year    = {2020},
  volume  = {216},
  pages   = {146},
  doi     = {10.1007/s11214-020-00771-x},
}

@article{Zhang2006,
  author  = {Zhang, M.},
  title   = {The theory of energetic particle transport in the magnetosphere: A noncanonical approach},
  journal = {Journal of Geophysical Research: Space Physics},
  year    = {2006},
  volume  = {111},
  pages   = {A04208},
  doi     = {10.1029/2005JA011149},
}

@article{Forbush_1937,
  author  = {Forbush, S. E.},
  title   = {On the Effects in Cosmic-Ray Intensity Observed During the Recent Magnetic Storm},
  journal = {Physical Review},
  year    = {1937},
  month   = jun,
  volume  = {51},
  number  = {12},
  pages   = {1108--1109},
  doi     = {10.1103/PhysRev.51.1108.3},
}

@article{Forbush_1938,
  author  = {Forbush, S. E.},
  title   = {On World-Wide Changes in Cosmic-Ray Intensity},
  journal = {Physical Review},
  year    = {1938},
  month   = dec,
  volume  = {54},
  number  = {12},
  pages   = {975--988},
  doi     = {10.1103/PhysRev.54.975},
}

@article{Forbush_1958,
  author  = {Forbush, Scott E.},
  title   = {Cosmic-Ray Intensity Variations during Two Solar Cycles},
  journal = {Journal of Geophysical Research},
  year    = {1958},
  month   = dec,
  volume  = {63},
  number  = {4},
  pages   = {651--669},
  doi     = {10.1029/JZ063i004p00651},
}

@article{Cane_2000,
  author  = {Cane, Hilary V.},
  title   = {Coronal Mass Ejections and {Forbush} Decreases},
  journal = {Space Science Reviews},
  year    = {2000},
  month   = jul,
  volume  = {93},
  pages   = {55--77},
  doi     = {10.1023/A:1026532125747},
}

@article{Richardson_2004,
  author  = {Richardson, Ian G.},
  title   = {Energetic Particles and Corotating Interaction Regions in the Solar Wind},
  journal = {Space Science Reviews},
  year    = {2004},
  month   = apr,
  volume  = {111},
  number  = {3},
  pages   = {267--376},
  doi     = {10.1023/B:SPAC.0000032689.52830.3e},
}

@article{Richardson_2011,
  author  = {Richardson, I. G. and Cane, H. V.},
  title   = {Galactic Cosmic Ray Intensity Response to Interplanetary Coronal Mass Ejections/Magnetic Clouds in 1995--2009},
  journal = {Solar Physics},
  year    = {2011},
  month   = jun,
  volume  = {270},
  number  = {2},
  pages   = {609--627},
  doi     = {10.1007/s11207-011-9774-x},
}

@article{Dumbovic_2012,
  author  = {Dumbovi{\'c}, M. and Vr{\v{s}}nak, B. and {\v{C}}alogovi{\'c}, J. and {\v{Z}}upan, R.},
  title   = {Cosmic ray modulation by different types of solar wind disturbances},
  journal = {Astronomy \& Astrophysics},
  year    = {2012},
  month   = feb,
  volume  = {538},
  pages   = {A28},
  doi     = {10.1051/0004-6361/201117710},
}

@article{Lara_etal_2024,
  author  = {Lara, Alejandro and Borgazzi, A. and Guennam, Eduardo and Niembro, Tatiana and Arunbabu, K. P.},
  title   = {Interaction of Cosmic Rays With Magnetic Flux Ropes},
  journal = {Journal of Geophysical Research: Space Physics},
  year    = {2024},
  volume  = {129},
  number  = {8},
  pages   = {e2024JA032478},
  doi     = {10.1029/2024JA032478},
}

\end{document}